\documentclass[aps,prl,twocolumn,groupedaddress,showpacs]{revtex4}
\usepackage[latin1]{inputenc}
\usepackage[english]{babel}
\usepackage{setspace}  
\usepackage{graphicx}
\usepackage{amssymb}
\usepackage{amscd}
\usepackage{rotating}
\usepackage{color}
\usepackage{epstopdf}

\begin{document}

\title{Thermal Canting of Spin-Bond Order}

\author{V.W.  Scarola$^{1,2}$, K.B. Whaley$^1$ and M. Troyer$^2$ }
\affiliation{$^1$Department of Chemistry and Pitzer Center for Theoretical Chemistry, University of California, Berkeley, California 94720, USA\\
$^2$Theoretische Physik, ETH Zurich, 8093 Zurich, Switzerland}

\date{\today}

\begin{abstract}

Magnetism arising from coupled spin and spatial degrees of freedom underlies the
properties of a broad array of physical systems.  We study here the interplay between correlations in spin and space for the quantum 
compass model in a finite external field, using quantum Monte Carlo methods.  We find that finite temperatures 
cant the spin and space (bond) correlations, with increasing temperature even reorienting spin correlations between 
orthogonal spatial directions.  We develop a coupled mean field theory to understand this effect in terms of 
the underlying quantum critical properties of crossed Ising chains in transverse fields and an effective field that weakens upon increasing temperature.  Thermal canting offers an experimental signature of spin-bond anisotropy.
\end{abstract}

\maketitle

\section{Introduction}
The collective behavior of spins in highly correlated quantum magnets typically establish  
order globally, over a large portion of the system by breaking a symmetry of the system.  Conventional 
characterizations of magnetic order, such as ferromagnetism, therefore rely on microscopic models 
with global symmetries.  Recent theoretical studies have, in contrast, analyzed models that support quantum phases without long-range ordering, for which novel types of order, including intriguing topological quantum liquid phases, can nevertheless arise \cite{wen1990, kitaev2003, kitaev2006}.
Quantum liquids can be difficult to identify (both theoretically and experimentally) because the lack of  long-range order defies 
characterization by simple bulk order parameters such as magnetization.  Models with quantum liquid ground states  
can harbor quasi-local symmetries which impose a symmetry on a small, non-extensive subset of spins \cite{nussinov2007}.  
Examples of such quasi-local operators include one-dimensional operator chains embedded in 
two-dimensional graphs.  The qualitatively distinct characteristics of systems with reduced, non-extensive symmetry
play an important role in identifying novel quantum liquid phases.  

We study here one of the simplest of these models, 
the quantum compass model in a magnetic field, $h$:
\begin{eqnarray}
  H=-\gamma \sum_{i,j}S^x_{i,j}S^x_{i+1,j}-  \sum_{i,j} S^z_{i,j} S^z_{i,j+1}-h\sum_{i,j} S^z_{i,j}, 
 \label{H-full}
\end{eqnarray}
where $S^{\alpha}_{i,j}=\sigma^{\alpha}_{i,j}/2$ are spin $1/2$ operators at the $(x,z)$ coordinates $(i,j)$ on 
an $L\times L$ square lattice.   It is sufficient to consider 
$\gamma>0$, since with a $\pi$ rotation of spins about the $z$-axis on a single sublattice, our results apply to the $\gamma<0$ case as well.   
The novel coupling between bond and spin degrees of freedom separates this class of models from more conventional models of 
magnetism and motivates the intriguing question as to whether there are generic signatures of anisotropic spin-bond coupling.

The quantum compass and related models were first discussed in the context of Mott insulators \cite{kugel1982}
and have been studied as simple models of orbital order in the transition metal oxide compounds with anisotropic coupling among orbitals defining pseudo-spins \cite{kugel1982,nussinov2004,tanaka2007}.  For example, recent experiments \cite{klingeler2005,senff2005,choi2008} on a 
two dimensional $e_g$ orbital 
compound, $\text{LaSrMnO}_4$, observe intriguing thermal effects.  Anomalies in thermal 
 expansion measurements \cite{klingeler2005} and in Raman scattering \cite{choi2008}, as well as structural changes measured by neutron and x-ray diffraction \cite{senff2005}, have all been interpreted as indicating a change of orbital occupation that is driven by temperature.
 The reorientation of orbital direction with 
 increasing temperature implied by the experiments raises the question as to whether this results from thermal repopulation of a non-interacting orbital system, or from competition between orbital-orbital interactions and thermal effects.
Detailed understanding of this phenomenon requires quantitative analysis of the competing roles of superexchange, phonon, crystal field and Jahn-Teller effects.   
However, the orbital-only degrees of freedom in these systems have been shown to be
 approximately described by a quantum compass model with anisotropic pseudo-spin interaction terms given by  superexchange contributions and an effective magnetic field determined by the crystal field splitting \cite{daghofer2006}.  This enables one to address in general the intriguing question of whether and how orbital-orbital interactions can influence the thermal redistribution of orbitals.  We answer this question in the affirmative here, showing that a novel type of thermal reorientation, 'thermal canting', of nearly degenerate orbitals can indeed arise from a large anisotropic orbital-orbital interaction of the type exemplified in Eq.~\ref{H-full}.   We also show that this thermal canting can provide an observable signature of the anisotropic interactions.  
 
Eq.~(\ref{H-full}) has been studied in other contexts as well.  
 A duality mapping \cite{nussinov2005} relates Eq.~(\ref{H-full}) to models of $p$-wave superconductors \cite{xu2004}.   
Recent proposals put forward to realize Eq.~(\ref{H-full}) in optical lattices, using 
asymmetric tunneling \cite{duan2003} or 
polar molecules \cite{zoller2006}, and in Josephson junction arrays \cite{doucot2005} (finite sized versions of which have recently been realized  \cite{gladchenko2008}), have been motivated by the recognition that the $h=0$ limit is characterized by sets of low energy 
two-fold degenerate states that may provide protected subspaces for encoding quantum information \cite{doucot2005,bacon2006}.
In addition, related \cite{nussinov2007} anisotropic spin models (Eq.~(4) in \cite{kitaev2006}) show topological order which allow 
a robust form of quantum information processing \cite{kitaev2003,nayak2007}.
A generic experimental signature of the spin-bond asymmetric models 
would be a valuable tool in the search for realizations of models with anisotropy and related quasi-local symmetries.   
a
In this paper we study the finite temperature properties of the quantum compass model for $h>0$, using both quantum Monte Carlo (QMC) and mean field theory to identify generic 
signatures of the novel set of one-dimensional symmetries underlying Eq.~(\ref{H-full}).  We find a remarkable thermal canting effect whereby finite temperatures \emph{enhance} spin-bond correlations $r_x$, where $r_{\alpha}= L^{-2}\sum_{i,j}\langle S^{\alpha}_{i,j}S^{\alpha}_{i+1,j} \rangle $ and $\alpha=x,z$.   The enhancement resembles an order-by-disorder process
 \cite{villian1980,nussinov2004,bergman2007}.   The reorientation of net spin-bond correlations (from $r_z$ to $r_x$) defines a phase akin to a liquid crystal smectic C phase where molecular moments order along a vector ('director') tilted with respect to translationally ordered columns.  We 
 develop a mean field theory using intersecting Ising models and their crossover phase diagrams \cite{sachdev1998} to show that the finite temperature reduction of the $z$-magnetization leads to thermal canting at the mean field level and should therefore play a role in similar models.  For $\gamma > 1$ we find that finite temperatures can completely reorient spin-bond correlations to point in orthogonal directions in spin and real space.  
 

\section{Symmetries in zero field limit}
We begin our study with an analysis of the unique symmetries of the quantum compass model in several well-defined
limits, for $h=0$.  Under 
high strain, $\gamma \gg1$ ($\gamma \ll 1$), the system forms a series of $L$ weakly coupled spin chains 
lying along the $x$ ($z$) direction in real space.  At zero temperature the chains order to point along the $x$ ($z$) directions in spin 
space.  Note that, with $h=0$, Eq.~(\ref{H-full}) commutes with the quasi-local symmetry operators
$
  P_i=\prod_j 2S^x_{i,j}
$ 
and
$
  Q_j=\prod_i 2S^z_{i,j},
$
which act on ordered chains to generate spin flips along $z$ and $x$ chains, respectively
($P$ is not an exact symmetry for $h\ne0$).  
Gapped excitations along chains are then finite 
length domains of flipped spins, as in one-dimensional Ising models.  The operators $Q$ ($P$) can flip all spins along the ordered chains in the ground state, resulting in yet another ground state. This results in a two-fold degeneracy of each chain and therefore a massive ground state 
degeneracy of the two-dimensional system of at least $2^L$.   

As $\gamma\rightarrow 1$, the 
inter-chain coupling allows a set of inter-chain excitations.  
Ref. \cite{dorier2005} showed that high order inter-chain fluctuations of the magnetization cost zero energy (for $L\rightarrow\infty$), preserving 
a large, $\mathcal{O}(2^L)$, ground state degeneracy for \emph{all} $\gamma$.  
The $\gamma=1$, $h=0$ limit yields discrete, global rotational symmetries  (see Ref.~\cite{nussinov2005} for a discussion). 
A nematic-like order parameter has been defined\cite{mishra2004,nussinov2005}: $r\equiv r_z-r_x$, to characterize smectic-like bond ordering along the $x$ and $z$ directions in spin and real space, analogous to the directional ordering of molecular moments along columns in liquid crystals.  $r = 0$ implies a disordered phase with no preferred direction.  Finite temperature studies \cite{tanaka2007,wenzel2008} suggest a phase transition from an ordered, $| r | >0$, phase to a disordered phase with $T_c\approx 0.055$  \cite{wenzel2008} in units of the interaction strength Eq.~(\ref{H-full}).   Example interaction 
strengths in corresponding experimental systems are $\sim 1000K$ for orbitals \cite{klingeler2005} or $\sim 1 K$ for Josephson junctions \cite{gladchenko2008}.  

\section{Quantum Monte Carlo simulations at finite fields}
A weak finite magnetic field picks out a single ground state with a gap to excitations set by $h$, simplifying QMC simulations.  We employ the stochastic series expansion using directed loops \cite{sandvik1999,sandvik2002,alet2005}.  A modification of the code in the ALPS \cite{ALPS} package allows a treatment of $S^x_{i,j}S^x_{i+1,j}$ bond terms in the updating scheme.   All QMC results presented here are converged to the thermodynamic limit in studies of several system sizes up to $L=30$ for periodic boundary conditions.

\begin{figure}[t] 
   \centering
   \includegraphics[width=3in,angle=0]{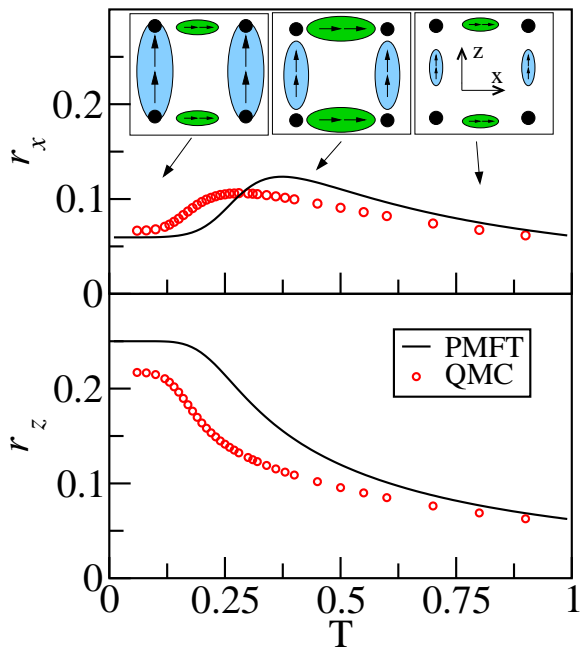} 
   \caption{The top (bottom) panels plot the spin-bond correlations versus temperature along the $x$ ($z$) direction computed via quantum Monte Carlo (open circles) and partial mean field theory (solid line) for the quantum compass model on a square lattice in a magnetic field.  The peak in $r_x$ indicates a reorientation of bond correlations with increasing temperature depicted schematically in the insert, where $\uparrow$ and $\rightarrow$ correspond to spins aligned along the $z$ and $x$ directions, respectively.  The data are plotted here for $\gamma=1$ and $h=0.08$.}
\label{fig1}
\end{figure}

Our results show that competition between the interaction and the magnetic field sets up an interesting 
interplay in the energetics along both spatial directions.  The insert in Fig.~\ref{fig1} (top panel) shows a schematic of 
the generic behavior observed in the QMC calculations for $\gamma \leq 1$ 
as we increase temperature.
At low temperatures the weak magnetic field aligns the net magnetization along the $z$-direction.  The magnetic field and $z$-interaction term favor spin chains that are bond-correlated along the $z$ direction, i.e. large $r_z$.  
Using mean field theory we will show that inter-chain interactions contribute to an
effective magnetic field that suppress excitations along the $x$-direction.  As we increase temperature above the gap set by $h$, the applied magnetic field becomes less relevant while the mean field effective magnetic field is reduced with increasing temperature.  Excitations along $x$ are then allowed to recover.  Consequently,  bond correlations build up along the $x$-direction yielding $r_x \rightarrow r_z$ as can be seen in the QMC and mean field results in Fig.~\ref{fig1}.  At high temperatures both types of correlations are trivially suppressed as the system becomes thermally disordered. 

\section{Partial Mean Field Theory}
At first sight, the build up of bond correlations along the $x$-direction is puzzling since one would expect a suppression of bond correlations with increasing temperature. 
To understand this build up of correlations along the $x$-direction, we construct a coupled 
partial mean field theory (PMFT) in which spin interactions along a given bond direction are treated at the mean field level while the 
interactions along the orthogonal direction are treated exactly.  In our two-step protocol we first choose the $z$ direction (indicated by 
$\langle ...\rangle_{z}$)
and make a mean field decoupling, $S^z_{i,j} S^z_{i,j+1}\rightarrow S^z_{i,j} \langle S^z_{i,j+1} \rangle_z$ to generate 
the effective field $B^{\text{eff}}_x=\langle S^z_{i,j+1} \rangle_{z}+\langle S^z_{i,j-1} \rangle_{z}+h$ for  spins along $x$.  
We then repeat the process by exchanging $x$ and $z$, resulting in a set of two coupled mean field equations:
\begin{eqnarray}
  H_{MF}^z&=&-  \sum_{i,j} S^z_{i,j} S^z_{i,j+1}-h\sum_{i,j} S^z_{i,j}-B^{\text{eff}}_z \sum_{i,j} S^x_{i,j}
 \label{H-MFTz}
 \\
 H_{MF}^x&=&-\gamma \sum_{i,j}S^x_{i,j}S^x_{i+1,j}- B^{\text{eff}}_x \sum_{i,j} S^z_{i,j}
 \label{H-MFTx}
\end{eqnarray}
While in general 
one may take $B^{\text{eff}}_z=\gamma(\langle S^x_{i+1,j}\rangle_{x}+\langle S^x_{i-1,j}\rangle_{x})$ and make a self-consistent solution to Eqs.~(\ref{H-MFTx}) and (\ref{H-MFTz}), in what follows we keep the mean field solutions analytic by setting $B^{\text{eff}}_z=0$.   This approximation does not qualitatively alter our results for $\gamma$ near unity.  We also assume a uniform net magnetization $\langle S^z \rangle_z= \langle S^z_{i,j} \rangle_z$.  The magnetization $\langle S^z \rangle_z$ obtained from solving Eq.~(\ref{H-MFTz}) can then be substituted directly into Eq.~(\ref{H-MFTx}).  This mean field approach is motivated by the work of Ref.~\cite{chen2007} but differs by i) having a set of two coupled mean field equations,  ii) addressing a new parameter regime of finite temperatures and magnetic fields, and iii) obtaining qualitatively accurate results without a self consistency loop.  The mean field approach is expected to be qualitatively accurate because the finite magnetic field suppresses  
fluctuations along the $z$ direction.   

\begin{figure}[t] 
   \centering
   \includegraphics[width=3in]{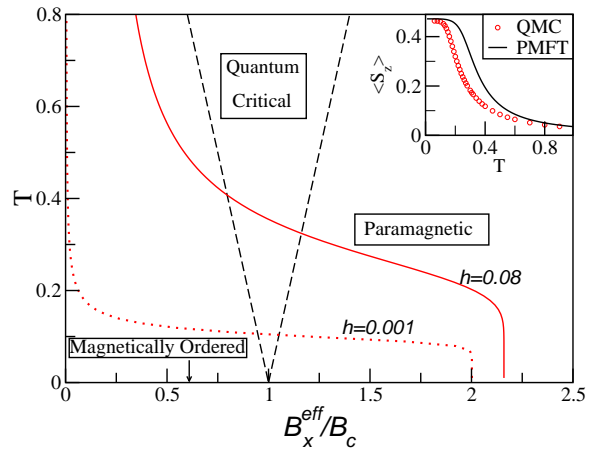} 
   \caption{Crossover diagram for a one-dimensional transverse field Ising model.  The $x$-axis indicates the effective magnetic field which can be lowered to tune from the paramagnetic side (weak nearest neighbor correlations), through a quantum critical point, $B_c\equiv \gamma/2$, to a regime with relatively strong nearest neighbor correlations.  The dashed lines schematically define the quantum critical regime.  For the PMFT defined by Eq.~(\ref{H-MFTx}), $B^{\text{eff}}_x$ is temperature dependent and plotted for $h=0.08$ (solid line) and $h=0.001$ (dotted line) with $\gamma=1$.  Inset: The same as Fig.~\ref{fig1} but for magnetization density.  The PMFT curve plots $\langle S^z \rangle_x$. }
   \label{fig2}
\end{figure}

We can now solve the PMFT equations analytically.  Solving Eq.~(\ref{H-MFTz}) using the well known solutions of the classical Ising model \cite{marsh1966}, we obtain $\langle S^z \rangle_z=\tanh(C)/(2G)$, where $G\equiv [1+(\exp(-4K)-1)\text{sech}^2(C)]^{1/2}$, $K\equiv1/4T$, and $C\equiv h/2T$ and can estimate $r_z$ as:
$
\langle S^z_{i,j}S^z_{i,j+1}\rangle_z=(1/4)\partial_K  \ln \{\exp(K)\cosh(C)(1+G)\}.  
$
This first estimate for the magnetization is then used as input to Eq.~(\ref{H-MFTx}), which is the Ising chain in a transverse field and can be solved exactly via the Jordan-Wigner transformation \cite{lieb1961}.  We obtain the total energy:  $\langle H_{MF}^x \rangle_x=\sum_{k}\epsilon_k (n_k-1/2)$, where
$\epsilon_k=[(B^{\text{eff}}_x)^2-\gamma B^{\text{eff}}_x \cos{(k)} +\gamma^2/4]^{1/2}$ and $n_k$ is the Fermi distribution function.  Our PMFT estimate for the magnetization of the two-dimensional system becomes:  $\langle S^z\rangle_x=\sum_{k}[2B^{\text{eff}}_x-\gamma \cos{(k)}](4\epsilon_k)^{-1}\tanh{(\epsilon_k/2T)} $.  We can also obtain an estimate for  $r_x$:
$
\langle S^x_{i,j}S^x_{i+1,j} \rangle_x=-(1/\gamma L)\langle H_{MF}^x \rangle_x-(1/\gamma)B^{\text{eff}}_x\langle S^z \rangle_x.
$
By setting $B^{\text{eff}}_z=0$ we ignore the influence of the magnetization along $x$ on $r_z$.  The result is an over-estimate for $r_z$ at low temperatures.  Fig.~ \ref{fig1} and the inset of Fig.~\ref{fig2} show a comparison between the PMFT and QMC results for the temperature dependence of $r_x$, $r_z$, and $\langle S^z \rangle$.  
It is evident that the PMFT captures the qualitative features of the QMC results.  

The success of the PMFT analysis implies that finite temperature excitations of the two-dimensional quantum compass model 
can be approximated by excitations along crossed 
(coupled) one-dimensional Ising models.  The peculiar enhancement of $r_x$ with temperature can now be understood from the crossover phase diagram of the 
transverse field Ising model, Eq.~(\ref{H-MFTx}).  Figure~\ref{fig2} plots a schematic of this crossover phase diagram.
In our mean field analysis, $B^{\text{eff}}_x$  carries a temperature dependence because the $z$-spins demagnetize with increasing temperature.  In Fig.~\ref{fig2} we superpose $B^{\text{eff}}_x(T)$ from a solution of the mean field equations, which shows that an increase in temperature lowers the effective magnetic field seen by spins along the $x$ direction.  The bond correlations along the $x$ direction are therefore enhanced because of a consequent drastic reduction of $\langle S^z \rangle_z$.  We conclude that the nonlinear temperature 
dependence in the effective magnetic field experienced by the spin chains lying along the $x$ direction is therefore responsible for the thermal canting of spin-bond correlations observed in QMC.  

\begin{figure}[t]
   \centering
   \includegraphics[width=3in,angle=0]{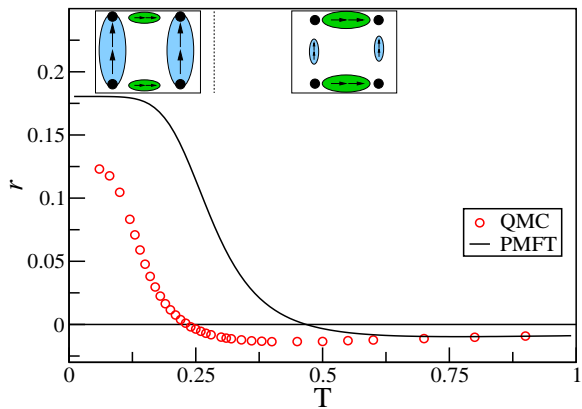} 
   \caption{Net spin-bond correlations as a function of temperature for $h=0.1$ and $\gamma=1/0.85$.  Thermal canting reorients the spin-bond correlations to lie predominantly along the $x$ direction in both spin and space for $T>0.25$.}
   \label{fig3}
\end{figure}

\section{Magnetic field dependence of thermal canting}
We now examine thermal canting in finite fields $h$ in greater detail.  Fig.~\ref{fig3} indicates that the net spin-bond correlations completely reorient with increasing temperature.  To understand this behavior, first note that 
the $h\sim \gamma \gg 1$ limit of Eq.~(\ref{H-full}) can be thought of as a set of nearly independent 
transverse field Ising models with a quantum critical point ($B_c=\gamma/2$).  Lowering $h$ takes the system from the paramagnetic phase through the quantum critical point (at $T=0$) to an ordered chain phase along the $x$ direction.  For a specific choice of $\gamma>1$ we can therefore completely reorient bond correlations by tuning $h$.  Thermal canting also tunes $B^{\text{eff}}_x$ but with a directionally independent parameter, temperature.  At low temperatures the finite magnetic field favors bond correlations along the $z$-direction, $r>0$, while at higher temperatures thermal canting of spin-bond correlations now favors a net orientation along the $x$-direction, $r<0$. This implies a canting transition at some temperature $T_{\text{cant}}$, which is illustrated in Fig.~\ref{fig3} for $\gamma=1/0.85$, where $T_{\text{cant}} \sim 0.25$.  These results predict a striking effect, namely, a complete reorientation of net bond correlations in both spin and real space with increasing temperature.  
This thermal reorientation arises from a distinct mechanism, thermal canting, that 
can again be understood in the PMFT calculations as being due to a weakening of the effective magnetic field with increasing temperature: thus it is a consequence of the two-dimensional, vector nature of the spin-bond correlations in  Eq.~(\ref{H-full}).  

\begin{figure}[t] 
   \centering
   \includegraphics[width=3in]{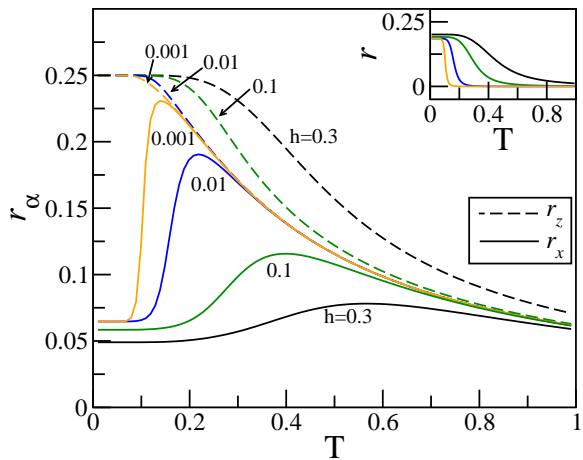} 
   \caption{PMFT results for the bond correlations along the $x$ (solid line) and $z$ (dashed line) directions versus temperature for $\gamma=1$ and several values of $h$.  The inset shows the behavior of the scalar difference $r=r_z-r_x$.}
   \label{fig4}
\end{figure}

Our two-dimensional spin-bond analysis at small finite $h$ also provides a microscopic rationale for the $\vert r \vert \rightarrow 0$ 
transition observed in the $h=0$ QMC simulations \cite{tanaka2007,wenzel2008} in terms of an 
immediate thermal canting of spin-bond correlations.  Our QMC and PMFT results both show that the $r_x$ peak in Fig.~\ref{fig1} moves to lower temperatures with decreasing magnetic field.  The mean field analysis indicates that this enhancement of $r_x$ at low temperatures derives from the rapid depolarization of $\langle S^z \rangle_z$ with increasing temperature when $h$ is small.  Mean field results for extremely small $h$ values and $\gamma=1$ are shown in Fig.~\ref{fig4}. 
We see that the increase in $r_x$  with temperature at low $h$ concomitantly tends to decrease $r$ (see inset).  
Extrapolating the 
PMFT results to $h\rightarrow 0$ yields, for $\gamma=1$, $r_x=(1/4)\tanh(1/4T)+\mathcal{O}(h^2)$ and $r=0+\mathcal{O}(h^2)$.  Thus 
$r$ disappears altogether for $h\rightarrow 0$.  This contrasts with the $h=0$ QMC results of Ref.~\cite{tanaka2007,wenzel2008} where a low temperature plateau in $\vert r \vert$ was followed by a rapid decrease at larger $T$ and assigned to a disordering transition.  The difference may be due to a lack of fluctuations in the PMFT which could play a large role in the $h\rightarrow 0$ limit.  

\section{Summary}
We have found a distinct, thermal canting effect in the $x$-$z$ plane of the two-dimensional quantum compass model.  Excitations governed by the rather unique set of chain symmetries implicit in the model provide a microscopic mechanism responsible for this effect.  
Thermal excitations above the paramagnetic gap can be approximated by those of intersecting Ising chains that are generated by quasi-local operators stemming from the  exact symmetry, $Q$, and the approximate symmetry, $P$.  Such excitations lead to an enhancement of spin-bond correlations along the $x$ direction at higher temperatures.  
This study shows that temperature can be used 
as a parameter to experimentally tune the balance between the excitations generated by these operators and to thereby reveal the underlying chain symmetries and anisotropic interactions.  Thermal canting arises in our mean field theory (as well as QMC).  Our partial
mean field argument applies to a number of other lattice geometries, in addition to the rectangular lattice of Eq.~\ref{H-full}.   

In summary, our combined QMC and partial mean field analysis show that anisotropic interactions can lead to a
thermal redistribution of spin-bond correlations. The results suggest that observation of thermal canting can provide an indicator of spin-bond anisotropy, with significant implications for the transition metal oxide compounds that are described by pseudo-spin models of orbital  degrees of freedom.  Our demonstration that thermal canting can be driven by anisotropic interactions suggests that the recent observations of a thermal reorientation of nearly degenerate orbitals in 
LaSrMnO$_4$ \cite{klingeler2005,senff2005,choi2008} warrant further analysis in terms of anisotropic pseudospin models of orbitals.

{\em Acknowledgements}  We thank NSF ITR and the Swiss NSF for support.  The simulations were performed on the ETH Brutus cluster.

\end{document}